\documentclass[11pt,twocolumn,twoside]{article}

\usepackage{fully3d}

\usepackage{acro}
\usepackage{bm}
\usepackage{amsfonts}

\usepackage{tikz}
\usetikzlibrary{arrows,shapes,intersections,calc,angles,decorations.pathreplacing,intersections,quotes,angles,positioning,fit,petri,through,shadows,plotmarks,spy}
\usepackage{pgfplots}
\usepackage{pgfplotstable}
\usetikzlibrary{pgfplots.groupplots}

\usepackage{algpseudocode}
\usepackage{algorithm}
\usepackage{graphicx}
\usepackage{caption,subcaption}

\usepackage{cancel}

\newcommand{\boldb}{\bm{b}}
\newcommand{\boldc}{\bm{c}}

\newcommand{\boldf}{\bm{f}}
\newcommand{\boldg}{\bm{g}}

\newcommand{\boldp}{\bm{p}}

\newcommand{\boldy}{\bm{y}}
\newcommand{\boldz}{\bm{z}}

\newcommand{\boldlambda}{\bm{\lambda}}
\newcommand{\boldmu}{\bm{\mu}}

\newcommand{\boldA}{\bm{A}}

\newcommand{\boldD}{\bm{D}}

\newcommand{\boldP}{\bm{P}}

\newcommand{\boldW}{\bm{W}}

\newcommand{\rme}{\mathrm{e}}

\newcommand{\R}{\mathbb{R}}
\newcommand{\transp}{^\top}

\newcommand{\ybar}{\bar{y}}

\newcommand{\gbar}{\bar{g}}

\newcommand{\boldmuhat}{\hat{\boldmu}}
\newcommand{\boldmutilde}{\tilde{\boldmu}}
\newcommand{\boldlambdahat}{\hat{\boldlambda}}
\newcommand{\boldlambdatilde}{\tilde{\boldlambda}}

\newcommand{\argmin}{\operatornamewithlimits{arg\,min}}

\algnewcommand{\Inputs}[1]{%
	\State \textbf{Inputs:}
	\Statex \hspace*{\algorithmicindent}\parbox[t]{.8\linewidth}{\raggedright #1}
}
\algnewcommand{\Initialize}[1]{%
	\State \textbf{Initialize:}
	\Statex \hspace*{\algorithmicindent}\parbox[t]{.8\linewidth}{\raggedright #1}
}

\DeclareAcronym{2D}{
	short=2-D,
	long=2-dimensional,
}

\DeclareAcronym{3D}{
	short=3-D,
	long=3-dimensional,
}

\DeclareAcronym{nD}{
	short=$n$-D,
	long=$n$-dimensional,
}

\DeclareAcronym{CT}{
	short=CT, 
	long=computed tomography,
}

\DeclareAcronym{PET}{
	short=PET, 
	long=positron emission tomography,
}

\DeclareAcronym{MRI}{
	short=MRI, 
	long=magnetic resonance imaging,
}

\DeclareAcronym{MR}{
	short=MR, 
	long=magnetic resonance,
}

\DeclareAcronym{PCCT}{
	short=PCCT, 
	long=photon-counting computed tomography,
}

\DeclareAcronym{DECT}{
	short=DECT, 
	long=dual-energy computed tomography,
}

\DeclareAcronym{MBIR}{
	short=MBIR, 
	long=model-based iterative reconstruction,
}

\DeclareAcronym{WLS}{
	short=WLS, 
	long=weighted least squares,
}
\DeclareAcronym{PWLS}{
	short=PWLS, 
	long=penalized weighted least squares,
}

\DeclareAcronym{PML}{
	short=PML, 
	long=penalized maximum likelihood,
}

\DeclareAcronym{CS}{
	short=CS,
	long=compressed sensing,
}

\DeclareAcronym{TV}{
	short=TV,
	long=total variation,
}

\DeclareAcronym{TNV}{
	short=TNV, 
	long=total nuclear variation,
}

\DeclareAcronym{JTV}{
	short=JTV, 
	long=joint total variation,
}

\DeclareAcronym{DTV}{
	short=DTV, 
	long=directional total variation,
}

\DeclareAcronym{PLS}{
	short=PLS, 
	long=parallel level sets,
}

\DeclareAcronym{SQS}{
	short=SQS, 
	long=separable quadratic surrogate,
}

\DeclareAcronym{ADMM}{
	short=ADMM, 
	long=alternating direction method of multipliers,
}

\DeclareAcronym{CDL}{
	short=CDL,
	long=convolutional dictionary learning,
}

\DeclareAcronym{MCDL}{
	short=CDL,
	long=multichannel convolutional dictionary learning,
}

\DeclareAcronym{CAOL}{
	short=CAOL, 
	long=convolutional analysis operator learning,
}

\DeclareAcronym{MCAOL}{
	short=MCAOL, 
	long=multichannel convolutional analysis operator learning,
}

\DeclareAcronym{PSNR}{
	short=PSNR, 
	long=peak signal-to-noise ratio,
}

\DeclareAcronym{SNR}{
	short=SNR, 
	long=signal-to-noise ratio,
}

\DeclareAcronym{CNN}{
	short=CNN, 
	long=convolutional neural network,
}

\DeclareAcronym{GAN}{
	short=GAN, 
	long=generative adversarial network,
}

\DeclareAcronym{VAE}{
	short=VAE, 
	long=variational autoencoder,
}

\DeclareAcronym{beta-VAE}{
	short=$\beta$-VAE, 
	long=beta-variational autoencoder,
}

\DeclareAcronym{LoR}{
	short=LoR,
	long=line of response,
	long-plural-form = lines of response,
}

\DeclareAcronym{MLEM}{
	short=MLEM,
	long=maximum likelihood expectation maximization,
}

\DeclareAcronym{GT}{
	short=GT,
	long=ground truth,
}

\begin{document}
%

\title{Synergistic PET/CT Reconstruction Using a Joint Generative Model} 

\author[1]{Noel Jeffrey Pinton}
\author[1]{Alexandre Bousse}
\author[1]{Zhihan Wang}
\author[1,2]{Catherine Cheze-Le-Rest}
\author[3]{Voichita Maxim}
\author[4]{Claude Comtat}
\author[4]{Florent Sureau}
\author[1]{Dimitris Visvikis}

\affil[1]{LaTIM, INSERM UMR 1101, \emph{Université de Bretagne Occidentale}, 29238 Brest, France.}
\affil[2]{Nuclear Medicine Department, Poitiers University Hospital, F-86022, Poitiers, France.}
\affil[3]{Université de Lyon, INSA‐Lyon, UCBL 1, UJM-Saint Etienne, CNRS, Inserm, CREATIS UMR 5220, U1294, F‐69621, LYON, France.}
\affil[4]{BioMaps, Université Paris-Saclay, CEA, CNRS, Inserm, SHFJ, 91401 Orsay, France.}

\maketitle
\thispagestyle{fancy}


\begin{customabstract}
We propose in this work a framework for synergistic \ac{PET}/\ac{CT} reconstruction using a joint generative model as a penalty. We use a synergistic penalty function that promotes \ac{PET}/\ac{CT} pairs that are likely to occur together. The synergistic penalty function is  based on a generative model, namely \ac{beta-VAE}. The model generates a \ac{PET}/\ac{CT} image pair from the same latent variable which contains the information  that is shared between the two modalities.

This sharing of inter-modal information can help reduce noise during reconstruction. Our result shows that our method was able to utilize the information between two modalities. The proposed method was able to outperform individually reconstructed images of \ac{PET} (i.e., by \ac{MLEM}) and \ac{CT} (i.e., by \ac{WLS}) in terms of \ac{PSNR}. Future work will focus on optimizing the parameters of the \ac{beta-VAE} network and further exploration of other generative network models.


\end{customabstract}

\section{Introduction}

\Ac{PET}/\ac{CT} hybrid imaging systems have been used since the 2000's. On one hand, \ac{PET} is an imaging modality used to observe and quantify molecular-level activities inside tissue through radioactive tracers, while on the other hand \ac{CT} is an imaging technique that uses X-rays to produce detailed structural images of the body. In order to get clearer images, both modalities use high dose of ionizing radiation which is detrimental to the health of certain patients especially children. Therefore reducing the dosage of used ionizing radiation is crucial, but it will affect the \ac{SNR} of the reconstructed images. 

Current multi-modal imaging reconstruction techniques process each modality separately. However, it is possible to exploit inter-modality information which could help reduce noise during reconstruction by combining functional \ac{PET} and structural \ac{CT} images. This approach could ultimately lead to reduction of dosage in patient imaging. 
The use of deep-learning in an unrolled \ac{MBIR} algorithm for synergistic \ac{PET}/\ac{MR} reconstruction \cite{corda2020synnet} has shown improved performance compared to independent conventional reconstruction methods. Although this approach is promising, the training is computationally expensive and the architecture depends on the scanners system projectors/back-projectors.


In this work, we propose a synergistic penalty function which uses generated \ac{PET} and \ac{CT} from the same latent variable which is trained on a \ac{beta-VAE}. As the shared information is contained in the penalty term, the proposed approach does not depend on the system (i.e. projector/back-projectors are not part of the training). 


Section~\ref{sec:method} describes our methodology and generative model, Section~\ref{sec:results} shows that out method performs better than traditional reconstruction techniques in terms of \ac{PSNR}, and in Section~\ref{sec:discussion} we discuss in what ways we can still improve the results.


\section{Method}\label{sec:method}

\subsection{Conventional PET/CT Reconstruction}

The linear X-ray attenuation and activity image are respectively denoted $\boldmu=[\mu_1,\dots,\mu_J]\transp \in \R^J_+$ and $\boldlambda=[\lambda_1,\dots,\lambda_J]\transp \in \R^J_+$, $J$ being be the number of pixels (or voxels) in the image. In the following we briefly introduce the \ac{MBIR} settings for \ac{CT} and \ac{PET}. 

We use a standard and monochromatic model for \ac{CT}. The transmission measurement are collected along $I$  rays and stored in a vector $\boldy = [y_1,\dots,y_I]\transp\in \mathbb{N}^I$. Given the attenuation $\boldmu$ and ignoring background events, the expected number of detected X-ray photons along each ray $i=1,\dots, I$ is given by the Beer-Lambert law as   
\begin{equation}\label{eq:beer-lambert}
	\ybar_i(\boldmu) = h \cdot \rme^{ -[\boldA \boldmu]_i}
\end{equation}	 
where $\boldA\in\R^{I\times J}$ is a discrete line integral operator such that $[\boldA]_{i,j}$ denotes the contribution of pixel $j$ to ray $i$, and $h$ is the X-ray intensity.  
Assuming that  the $I$ measurements $\{y_i\}$ are independent and follow a Poisson distribution centered on $\{\ybar_i(\mu)\}$, \ac{MBIR} of the attenuation image $\boldmu$ can be achieved by solving the following optimization problem with an iterative algorithm \cite{elbakri2002statistical}:
\begin{equation}\label{eq:reco_pwls}
	\boldmuhat \in \argmin_{\boldmu\ge\bm{0}} \, L_{\mathrm{CT}}(\boldmu)+ \alpha R_{\mathrm{CT}}(\boldmu)
\end{equation}
where the loss function $L_{\mathrm{CT}}$ is defined ad $L_{\mathrm{CT}} (\boldmu) = \frac{1}{2}\|\boldA \boldmu - \boldb\|^2_{\boldW} $ with $\boldb = -\log \boldy/ h \in \R^I$ being the vector of approximated line integrals and $\boldW = \mathrm{diag}\{\boldy\} \in \mathbb{N}^{I\times I}$ being a diagonal matrix of statistical weights,  $R_{\mathrm{CT}}$ is a penalty function or regularizer, generally convex, that promotes desired image properties (in general piece-wise smoothness) while controlling the noise, and $\alpha>0$ is a weight.

The \ac{PET} data are collected along $K$ \acp{LoR} defined by $K$ pairs of detectors. The emission data are stored in a vector $\boldg\in\mathbb{N}^K$. Given the activity image $\boldlambda$ and again ignoring background events, the expected number of detected $\gamma$-photon pairs along \ac{LoR} $K$ is
\begin{equation}\label{eq:pet_model}
	\gbar_k(\boldlambda) = \tau[\boldP \boldlambda]_k
\end{equation}	 
where $\boldP\in\R^{K\times J}$ is the \ac{PET} system matrix, i.e., $[\boldP]_{k,j}$ is the probability that a emission at pixel $j$ is detected along \ac{LoR} $k$ (accounting for attenuation), and $\tau$ is the acquisition time. Assuming that  the $K$ measurements $\{g_k\}$ are independent and follow a Poisson distribution centered on $\{\gbar_k(\boldlambda)\}$, \ac{MBIR} of the activity image can be achieved by solving the following \ac{PML} problem iteratively:
\begin{equation}\label{eq:reco_pet}
	\boldlambdahat \in \argmin_{\boldlambda\ge\bm{0}} \, L_{\mathrm{PET}}(\boldlambda) + \delta R_{\mathrm{PET}}(\boldlambda)
\end{equation}
where the loss function $L_{\mathrm{PET}}$ is the negative Poisson  log-likelihood defined as $L_{\mathrm{PET}}(\boldlambda)= \sum_{k=1}^K - g_k \log \gbar_k(\boldlambda) + \gbar_k(\boldlambda)$  and $R_\mathrm{PET}$ and $\delta$ are same as $R_\mathrm{CT}$ and $\alpha$ in  \eqref{eq:reco_pwls}.

\subsection{Synergistic PET/CT Reconstruction Using a Joint Generative Model}

$\boldmu$ and $\boldlambda$ can be simultaneously reconstructed by solving the following joint estimation problem:
\begin{equation}\label{eq:syn}
	(\boldmuhat,\boldlambdahat) \in \argmin_{\boldmu,\boldlambda\ge\bm{0}} \, L_{\mathrm{CT}}(\boldmu)+L_{\mathrm{PET}}(\boldlambda) +  \gamma R_{\mathrm{syn}}(\boldmu,\boldlambda)
\end{equation}
where $\gamma>0$ and $R_{\mathrm{syn}}$ is a \emph{synergistic} penalty function that promotes structural and/or functional correlations between the multiple images, such as parallel levelsets  \cite{ehrhardt2014joint} and \acl{TNV}  \cite{rigie2015joint}.  

Instead of using a handcrafted synergistic penalty function, $R_{\mathrm{syn}}$ may be trained such that $R_{\mathrm{syn}}(\boldmu,\boldlambda) \approx 0$ if  $\boldmu$ and $\boldlambda$ are images that are       plausible not only individually, but also    \emph{together}. In this work we propose the following penalty inspired by previous work from \citeauthor{duff2021regularising} \cite{duff2021regularising}:
\begin{align}
	R_{\mathrm{syn}}(\boldlambda,\boldmu)  & =   \min_{\boldz\in\R^P} \, \eta \frac{1}{2} \| \boldf_{\mathrm{PET}}(\boldz) - \boldlambda  \|^2 \nonumber \\
	& +  (1-\eta) \frac{1}{2} \| \boldf_{\mathrm{CT}}(\boldz) - \boldmu  \|^2  \label{eq:syn_penalty}
\end{align}
where $\boldf_{\mathrm{CT}}$ and $\boldf_{\mathrm{PET}}$ are the decoders of the trained generative model that generate \ac{CT} and \ac{PET} images from the same latent variable $\boldz \in \R^P$ and $\eta\in[0,1]$. To summarize \eqref{eq:syn_penalty}, $R_{\mathrm{syn}}(\boldmu,\boldlambda) \approx 0$ if $\boldmu$ and $\boldlambda$ are approximately generated from the same latent variable $\boldz$. This approach is somehow a generalization of coupled dictionary learning for multi-modal imaging (see for example  \cite{sudarshan2020joint}) where we replaced the dictionaries by generative models. Note that a penalty can be added  in  \eqref{eq:syn_penalty} to control the noise. In this work we used \acp{beta-VAE} for $\boldf_{\mathrm{CT}}$ and $\boldf_{\mathrm{PET}}$.

The overall reconstruction method is described in Algorithm~\ref{alg:alg}. The activity and attenuation images were initialized using standard \ac{MBIR} methods (\ac{MLEM} and \ac{WLS}). The algorithm then alternates between minimization in $\boldz$ (L-BFGS \cite{zhu1997algorithm}) , $\boldlambda$ (modified \ac{MLEM} \cite{depierro1995}) and $\boldmu$ (\ac{PWLS} \cite{elbakri2002statistical}). 

\begin{algorithm}[ht]
	\caption{Synergistic reconstruction of PET/CT images 
		}

		\label{alg:alg}
		\textbf{Input:} Maximum iteration number \texttt{MaxIter}, sub-iteration numbers \texttt{SubIter1} and \texttt{SubIter2}, penalty parameter $\gamma$, modal parameter $\eta$
		\begin{algorithmic}[1]
			\Inputs{$\gamma$, $\eta$, $\boldg$, $\boldy$}
			\Initialize{$\boldlambda^0 = \boldlambda^{\mathrm{init}}$, $\boldmu^0=\boldmu^\mathrm{init}$, $\boldz^0=\bm{0}$; $\boldp = \boldP\transp \bm{1}$; $\boldb = -\log \boldy/ h \in \R^I$ }
			
			\For{$n = 1$ to \texttt{MaxIter}}
			\State $\boldz^n=\argmin_{\boldz\in\R^P} \eta \frac{1}{2} \|\boldf_{\mathrm{PET}}(\boldz)-\boldlambda^{n-1}\|^2 + (1-\eta)\frac{1}{2}\|\boldf_{\mathrm{CT}}(\boldz)-\boldmu^{n-1}\|^2$   \Comment{\emph{using LBFGS}}
			\State $\boldlambdatilde \leftarrow \boldlambda^{n-1} $ 
			\For{$l = 1$ to $\texttt{SubIter1}$}
			\State $\boldlambdatilde^{\mathrm{em}} \leftarrow \boldlambdatilde \frac{1}{\boldp} \boldP\transp \left[\frac{\boldg}{\boldP\boldlambdatilde}  \right]$
			\State $\boldc \leftarrow \gamma\eta \boldf_{\mathrm{PET}}(\boldz^n) - \boldp$
			\State $\boldlambdatilde \leftarrow \frac{\boldc + \sqrt{\boldc^2 + 
					4\gamma\eta \boldp \boldlambda^{\mathrm{em}}}}{2\gamma\eta}$
			\EndFor
			\State $\boldlambda^n = \boldlambdatilde $
			
			\State $\boldmutilde \leftarrow \boldmu^{n-1}$
			
			\For{$m = 1$ to $\texttt{SubIter2}$}
			\State $\boldmutilde^{\mathrm{rec}} \leftarrow  \boldmutilde - \boldD^{-1} \boldA\transp \boldW (\boldA \boldmutilde - \boldb)  $
			\State $\boldmutilde \leftarrow \left[\frac{\boldD \boldmutilde^{\mathrm{rec}} + \gamma (1-\eta) \boldf_{\mathrm{CT}}(\boldz^n) }{\boldD + \gamma (1-\eta)}\right]_+$
			\EndFor
			\State $\boldmu^n =  \boldmutilde $
			
			\EndFor
			
			\State \Return $(\boldlambda^\texttt{MaxIter}, \boldmu^\texttt{MaxIter})$
		\end{algorithmic}
	\end{algorithm}

\subsection{Generative Model Network Architecture}

\begin{figure}[!ht]
	\centering
	
	\includegraphics[width=\linewidth]{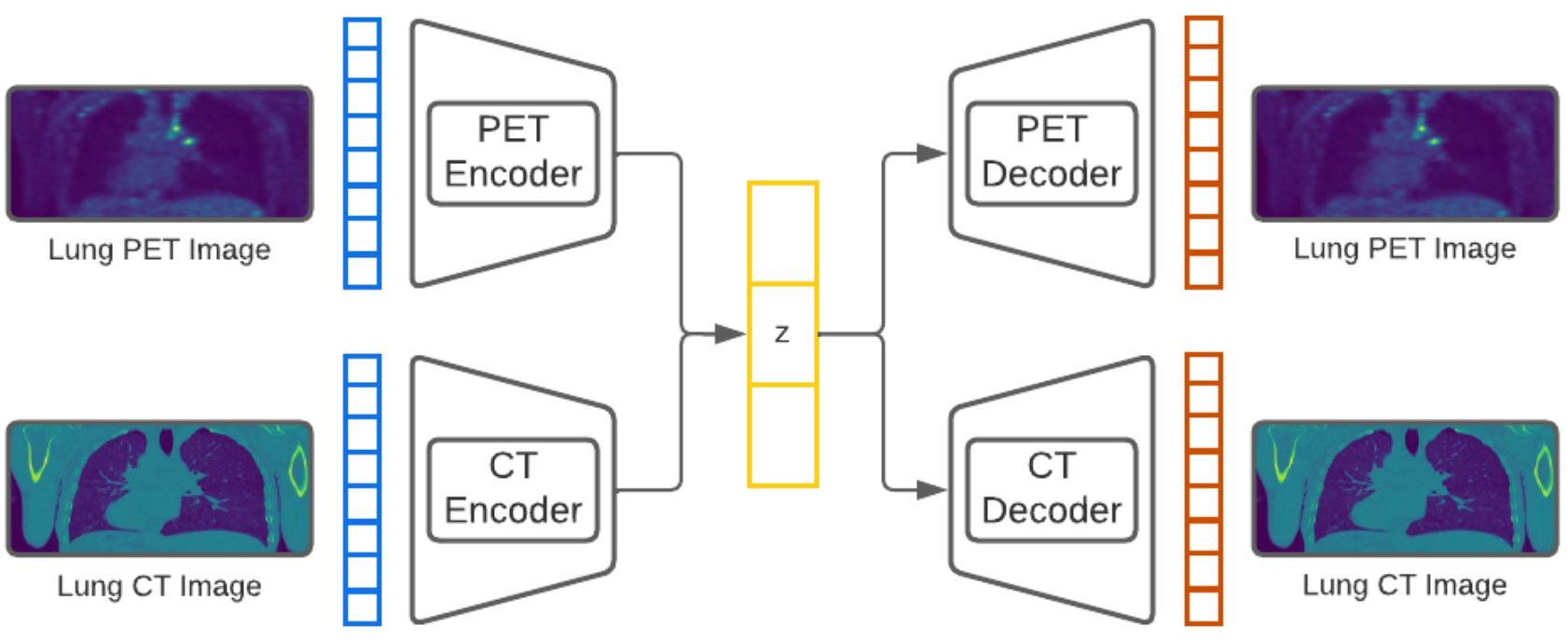}
	
	\caption{Multi-branch $\beta$-VAE architecture}
	\label{fig:VAEArchitecture}
\end{figure}

For combining functional \ac{PET} and structural \ac{CT} images, we created a multi-branch \ac{beta-VAE} (Figure \ref{fig:VAEArchitecture}) architecture. The network features separate encoders and decoders for \ac{PET} and \ac{CT}. The encoders and decoders consist 5-layer networks with ReLU activation. We train the networks using the following loss function as described in \cite{higgins2017betavae}:
\begin{align}\label{eq:loss_betavae}
	\mathrm{loss} & = \mathbb{E}_{q_\phi(\boldz|\boldlambda, \boldmu)}[\log p_\theta(\boldlambda, \boldmu \mid \boldz)]
	\nonumber \\
	& - \beta D_\mathrm{KL} (q_\phi (\boldz|\boldlambda, \boldmu)\,\, \|\,\, p(\boldz))
\end{align}
where $\beta$ is the regularization coefficient that constrains the capacity of the latent information channel $\boldz$. The first term is the log-likelihood of the observed data $p_\theta(\boldlambda, \boldmu | \boldz)$  averaged over the  latent variable $\boldz$ with distribution $q_\phi(\boldz|\boldlambda, \boldmu)$, also referred to as the cross-entropy. The second term is the Kullback-Leibler divergence between the posterior variational approximation $q_\phi (\boldz|\boldlambda, \boldmu)$ and the prior distribution $p(\boldz)$, which is selected to be a standard Gaussian distribution. 

\subsection{Network Training and Implementation}

The models were trained using a collection of \ac{PET}/\ac{CT} lung image pair $(\boldlambda, \boldmu)$ of 117 patients from CHU Poitiers, Poitiers, France. The images are $256\times256\times 128$, each of the 128 slice pairs taken independently, resulting in $117\times 128 = 14,976$  \ac{PET}/\ac{CT} image pairs for training. The \ac{PET} images were clipped at $10^5$ Bq/mL in order to remove outlier. Both \ac{PET} and \ac{CT} images were scaled between 0 and 1 for training. During the final reconstruction, the \ac{PET} and \ac{CT} images are returned to their original domain values. A fixed value of $\beta$ at $100$ of the loss function (\ref{eq:loss_betavae}) was used for this study. We used Tensorflow v2.4 and Keras for the network implementation and training. The Adam Algorithm \cite{adam2014} was used with a learning rate of 0.0001 for 300 epochs and a batch size of 128 on a NVIDIA RTX A6000 GPU. 
The gradients were computed using the \texttt{GradientTape} function from Tensorflow for the L-BFGS optimization \cite{zhu1997algorithm} used in the minimization over $\boldz$ in \eqref{eq:syn_penalty}, which uses “automatic differentiation”, breaking complex gradient calculation into simpler gradient calculations through chain rules.

\subsection{Raw Data Generation}

For the quantitative analysis of the reconstruction algorithm, we generated low-count  \ac{PET}/\ac{CT} raw data as
\begin{align}
	g_k & \sim \mathrm{Poisson}(\gbar_k(\boldlambda^\star))   \label{eq:gk}  \\
	y_i & \sim \mathrm{Poisson}(\ybar_i(\boldmu^\star))	\label{eq:yi}
\end{align}
where $\boldlambda^\star$ and $\boldmu^\star$ are respectively \ac{PET} and \ac{CT} images that we used as a \ac{GT}. Attenuation was ignored for the \ac{PET} data. 
We used SciPy \cite{2020SciPy-NMeth} for the \ac{PET} projectors (Radon transform) and Astra Toolbox \cite{vanAarle:16} for \ac{CT} projectors (for both data generation and reconstruction).


\section{Results}\label{sec:results}

\subsection{Network-based Penalty Effect}\label{subsec:networkeffect}

The parameter $\gamma$ in Equation~\eqref{eq:syn} determines the penalty function contribution. Higher $\gamma$ values pushes the algorithm to favor the generated image from the neural network than the data fidelity term.
In order to quantitatively assess the efficacy of inter-modality information sharing in the synergistic reconstruction of \ac{PET} and \ac{CT}, we reconstructed the \ac{PET} and \ac{CT} images with different values of $\gamma$ (cf. Equation~\eqref{eq:syn}), starting from $\gamma=0$ which corresponds to standard independent reconstructions, i.e., \ac{MLEM} for \ac{PET} and \ac{WLS} for \ac{CT}. 

\begin{figure}
	\centering
	\captionsetup[subfigure]{justification=centering}
	\begin{subfigure}[t]{0.15\textwidth}
		\includegraphics[width=\linewidth]{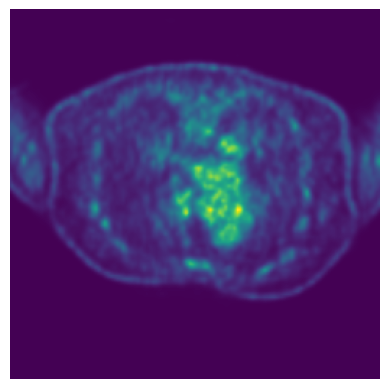}
		\caption*{Ground truth}
	\end{subfigure}
	\begin{subfigure}[t]{0.15\textwidth}
		\includegraphics[width=\linewidth]{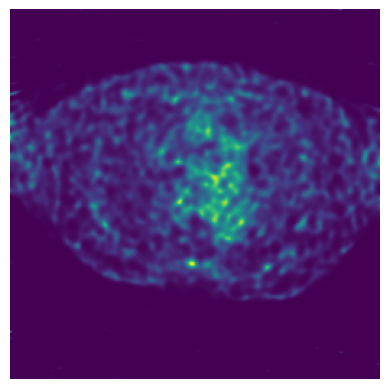}
		\caption*{\ac{MLEM} (PSNR$=25.06$)}
	\end{subfigure}
	\begin{subfigure}[t]{0.15\textwidth}
		\includegraphics[width=\linewidth]{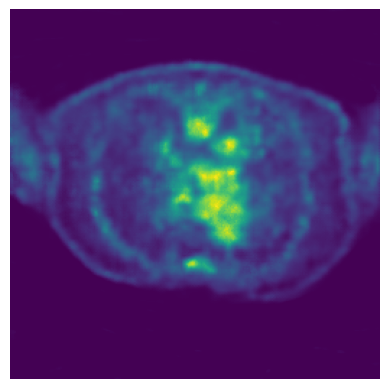}
		\caption*{Our Method (PSNR$=27.96$)}
	\end{subfigure}
	
	\begin{subfigure}[t]{0.15\textwidth}
		\includegraphics[width=\linewidth]{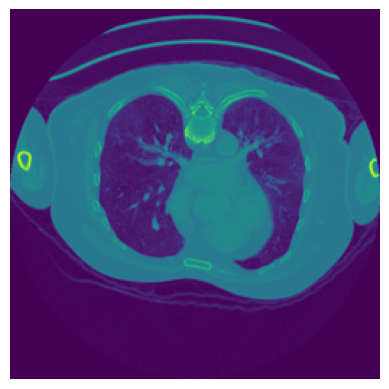}
		\caption*{Ground truth}
	\end{subfigure}
	\begin{subfigure}[t]{0.15\textwidth}
		\includegraphics[width=\linewidth]{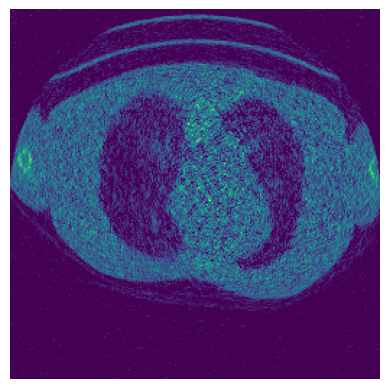}
		\caption*{\ac{WLS} \\ (PSNR$=19.78$)}
	\end{subfigure}
	\begin{subfigure}[t]{0.15\textwidth}
		\includegraphics[width=\linewidth]{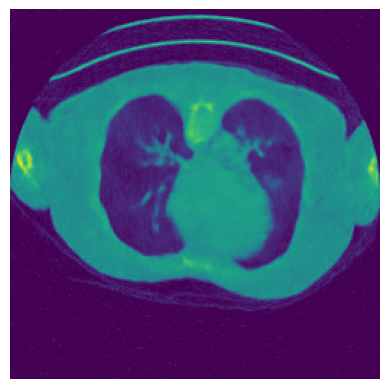}
		\caption*{Our Method (PSNR$=30.53$)}
	\end{subfigure}
	\caption{Comparison of independent \ac{PET} and \ac{CT} reconstructions with synergistic \ac{PET}/\ac{CT} reconstruction using our method at $\gamma=5\times10^5$ and $\eta=0.5$. 
	}
	\label{fig:petct_images}
	
\end{figure}

Figure~\ref{fig:petct_images} shows the reconstruction of \ac{PET} (\ac{MLEM}) and \ac{CT} (\ac{WLS}) independently (no penalties) and the synergistic reconstruction of the \ac{PET}/\ac{CT} image pair (our method).



\begin{figure}[ht]
	\centering
	\begin{tikzpicture}[scale=0.5] 
	
	\begin{axis}[
	mark options={mark size = 3pt},
	xlabel={$\gamma$},
	ylabel={PSNR},
	xmin = 0,
	xmax = 1000000,
	legend cell align=left,
	legend style={at={(0.7,0.3)},anchor=west},
	x post scale=2,
	]
	
	\addplot[color=red, style={thick}, mark=square*] table[x=beta, y=psnr, col sep=comma] {./data/pet_gamma/pet_gamma0.5.txt};
	
	\addplot[color=blue, style={thick}, mark=square*] table[x=beta, y=psnr, col sep=comma] {./data/ct_gamma/ct_gamma0.5.txt};
	
	\addlegendentry{PET}
	\addlegendentry{CT}
	
	\end{axis}

	\end{tikzpicture}
	\caption{PSNR values of reconstructed PET/CT images with respect to $\gamma$ at $\eta=0.5$.}
	\label{fig:PSNRvsGamma}
\end{figure}
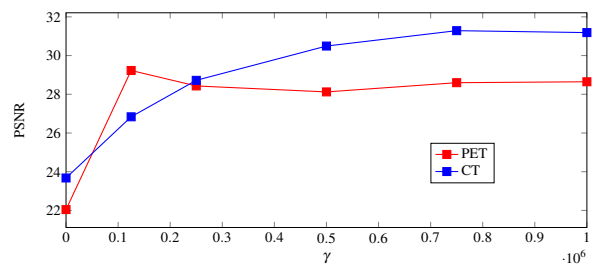

The final reconstructed images by our method show that the synergistic reconstruction of \ac{PET} and \ac{CT} with the help of generative models as a penalty term performs better in terms of \ac{PSNR} than individually reconstructed \ac{PET} and \ac{CT} images. We can infer from this result that both \ac{PET} and \ac{CT} are benefiting from the inter-modality information from each other. 
Figure~\ref{fig:PSNRvsGamma} shows the graph of \ac{PSNR} with respect to $\gamma$ for both \ac{PET} and \ac{CT}. We can see from the graph that an increase in the parameter $\gamma$ corresponds to an increase in \ac{PSNR} for \ac{CT}. For \ac{PET}, the \ac{PSNR} reaches a maximum and then starts to decline to a stable value. We observe that the optimum was not reached for the same $\gamma$. A solution is proposed on the next section.

\subsection{Inter-modality Information Effect}\label{subsec:intermodal}

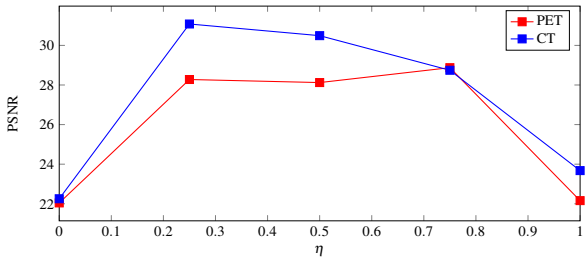
\begin{figure}
	\centering
	\begin{tikzpicture}[scale=0.5] 
	
	\begin{axis}[
	mark options={mark size = 3pt},
	xlabel={$\eta$},
	ylabel={PSNR},
	xmin = 0,
	xmax = 1,
	legend cell align=left,
	x post scale=2,
	]
	
	\addplot[color=red, style={thick}, mark=square*] table[x=eta, y=psnr, col sep=comma] {./data/pet_eta/pet_eta500000.0.txt};
	\addlegendentry{PET}
	\addplot[color=blue, style={thick}, mark=square*] table[x=eta, y=psnr, col sep=comma] {./data/ct_eta/ct_eta500000.0.txt};
	\addlegendentry{CT}
	\end{axis}

	\end{tikzpicture}
	\caption{PSNR values of reconstructed PET image with respect to $\eta$ at $\gamma=5\times10^5$.}
	\label{fig:PSNRvsEta}
\end{figure}

Figure~\ref{fig:PSNRvsEta} shows the effect of $\eta$  from Equation (\ref{eq:syn_penalty}). The variable $\eta$ is a weight that represents the contribution of \ac{PET} and \ac{CT} generated image into the final reconstruction. A value of $\eta=0$ corresponds to full contribution from CT generated image and no contribution from  PET in the prior, and vice versa. As shown in the graph (Figure \ref{fig:PSNRvsEta}), both curves have a maximum somewhere between $\eta=0$ and $\eta=1$. This shows that the reconstructed \ac{PET} and \ac{CT} images were able to exploit the inter-modal information. However, we observe that the optimum was not reached for the same values of $\eta$, the same as the previous section with $\gamma$. 
We conjecture that this could be fixed by weighting the data fidelity terms $L_{\mathrm{CT}}$ and  $L_{\mathrm{PET}}$ differently as seen in Equation~\ref{eq:weighteddatafidelity},
\begin{align}\label{eq:weighteddatafidelity}
	(\boldmuhat,\boldlambdahat) \in \argmin_{\boldmu,\boldlambda\ge\boldsymbol{0}} \, \rho L_{\mathrm{CT}}(\boldmu) &+ (1-\rho) L_{\mathrm{PET}}(\boldlambda) \nonumber \\ &+  \gamma R_{\mathrm{syn}}(\boldmu,\boldlambda) \, .
\end{align}

\section{Discussion}\label{sec:discussion}

In this paper, we proposed a synergistic \ac{PET}/\ac{CT} reconstruction which utilizes a deep penalty method using a generative neural network \ac{beta-VAE}. The results showed that the technique outperforms traditional \ac{PET}-only and \ac{CT}-only reconstruction methods for both modalities. We have also demonstrated that the \ac{PET} and \ac{CT} images were able to exploit the inter-modality information between each other. The use of \ac{beta-VAE} generators were effective, but fine tuning the hyperparameters are needed in order to produce clearer images. Other suggestions include the tuning of the $\beta$ variable in Equation (\ref{eq:loss_betavae}) and the use of attention modules in the latent space. The possibility of using other generative models such as \ac{GAN}-based models and diffusion-based models is also open. Furthermore, the addition of weights in the first two terms of Equation \ref{eq:syn} may improve the results by optimizing the contribution of each term to the final output. Finally, we ignored attenuation factors in the \ac{PET} reconstruction. In principle they should be computed from a first reconstruction of the attenuation (non synergistic). Since they correspond to line integrals, they do not suffer from noise amplification.
\section{Conclusion}

We proposed in this work a framework for synergistic \ac{PET}/\ac{CT} reconstruction using a generative neural network as penalty. The decoders of the trained beta-\ac{VAE} were able to effectively promote correlations between the two modalities. Evaluations using real patient dataset indicated that the proposed framework was able to exploit the inter-modal information between the two modalities  which ultimately led to improved image quality.
 
\section*{Acknowledgement}
This work was funded by the French National Research Agency (ANR) under grant ANR-20-CE45-0020.

\printbibliography

\end{document}